\documentclass[prl,twocolumn,showpacs]{revtex4}
\usepackage[dvips]{graphicx}

\newcommand{\ignore}[1]{}

\begin{document}

\title
{Intrinsic effects of the boundary condition on switching processes in spin-crossover solids}

\author{Masamichi Nishino$^{1,2,6}$}  
\email[Corresponding author. Email address: ]{nishino.masamichi@nims.go.jp} 
\author{Cristian Enachescu$^{3}$}
\author{Seiji Miyashita$^{4,6}$}
\author{Kamel Boukheddaden$^{5}$}
\author{Fran\c cois Varret$^{5}$}
\affiliation{$^{1}${\it Computational Materials Science Center, National Institute}  
for Materials Science, Tsukuba, Ibaraki 305-0047, Japan \\
$^{2}${\it Department of Theoretical and Computational Molecular Science, 
Institute for Molecular Science, Okazaki, Japan} \\
$^{3}${Department of Physics, Alexandru Ioan Cuza University, Iasi, Romania} \\$^{4}${\it Department of Physics, Graduate School of Science,} The University of Tokyo, Bunkyo-Ku, Tokyo, Japan  \\
$^{5}${\it  Groupe d'Etudes de la Mati\`{e}re Condens\'{e}e, CNRS-Universit\'{e}} 
de Versailles/St. Quentin en Yvelines, 45 Avenue des Etats Unis, F78035 Versailles Cedex, France \\
$^{6}${\it CREST, JST, 4-1-8 Honcho Kawaguchi, Saitama, 332-0012, Japan}
}
\date{\today}

\begin{abstract}

We investigated domain growth in switching processes between the low-spin and high-spin phases in thermally induced hysteresis loops of spin-crossover (SC) solids. 
Elastic interactions among the molecules induce 
effective long-range interactions, and thus the boundary condition 
plays a significant role in the dynamics. 
In contrast to SC systems with periodic boundary conditions, 
where uniform configurations are maintained during the switching process, 
we found that domain structures appear with open boundary conditions. 
Unlike Ising-like models with short-range interactions, domains always grow from the corners of the system. The present clustering mechanism provides an insight into the switching dynamics of SC solids, in particular, in nano-scale systems.

\end{abstract}

\pacs{75.30.Wx 75.50.Xx 75.60.-d 64.60.-i}

\maketitle

Spin-crossover (SC) compounds have been studied intensively because of their peculiar physical properties due to competition between the low energy of the low-spin (LS) state and the high entropy of the high-spin (HS) state \cite{Gutlich_book,Konig,Hauser,Sorai}.  
SC transitions are induced by changes in temperature, pressure, etc. 
The LS state can be excited by photo-irradiation to a long-lived HS state at low temperatures, which is called LIESST (light induced excited spin state trapping) \cite{DECURTINS}, and reverse LIESST (HS to LS) can also be 
obtained at a different wavelength \cite{Hauser2}. 
These controllable and functional properties \cite{Gutlich_book,Tayagaki,Letard,Gawelda,Lorenc,Fouche} would bring potential applicability to novel optical devices, e.g., optical data storage and optical sensors.

The LS and HS states couple through a vibronic mechanism and the size of 
the SC molecule changes with the spin state. 
The distortion caused by the change of molecular size induces 
a kind of elastic interaction among the spin states of molecules. 
The importance of the elastic interaction has been reported for SC transitions \cite{Gutlich_book,Zimmermann,Adler,Willenbacher,Nishino_elastic,Miyashita_elastic,Konishi_elastic,Kamel_elastic,Niclazzi,Nishino_elastic2,Cristian_elastic}. 
The mechanism of the phase transition induced by the elastic interaction has been studied and various new aspects have been revealed.
For example, using periodic boundary conditions, 
it has been shown that effective long-range interactions suppress 
domain growth, and uniform configurations are maintained even near the critical temperature \cite{Miyashita_elastic}.  
In the process of switching the configuration uniformity is also maintained, 
which is considered an intrinsic property of systems with the effective long-range interactions. 
In these systems, we do not expect the critical opalescence due to growth of large clusters.

Nowadays SC compounds are a focus of nano-science and technology \cite{Cobo,Coronado,Molnar,Volatron, Boldog}. 
On the nano-scale, such as powder or thin film samples, 
the boundary effect is important. In particular, in systems with long-range interactions the concept of the thermodynamic limit may not be well defined and the effect of the boundary must be considered carefully.  

In this Letter we investigate how a SC system with effective long-range interactions switches between the bistable states, using open boundary conditions. We analyze characteristic features of the heating and cooling processes 
in thermal hysteresis loops with open boundary conditions (OBC), and 
compare to periodic boundary conditions (PBC).

\begin{figure}
 \begin{minipage}{0.49\hsize}
  \begin{center}
     \includegraphics[width=41mm]{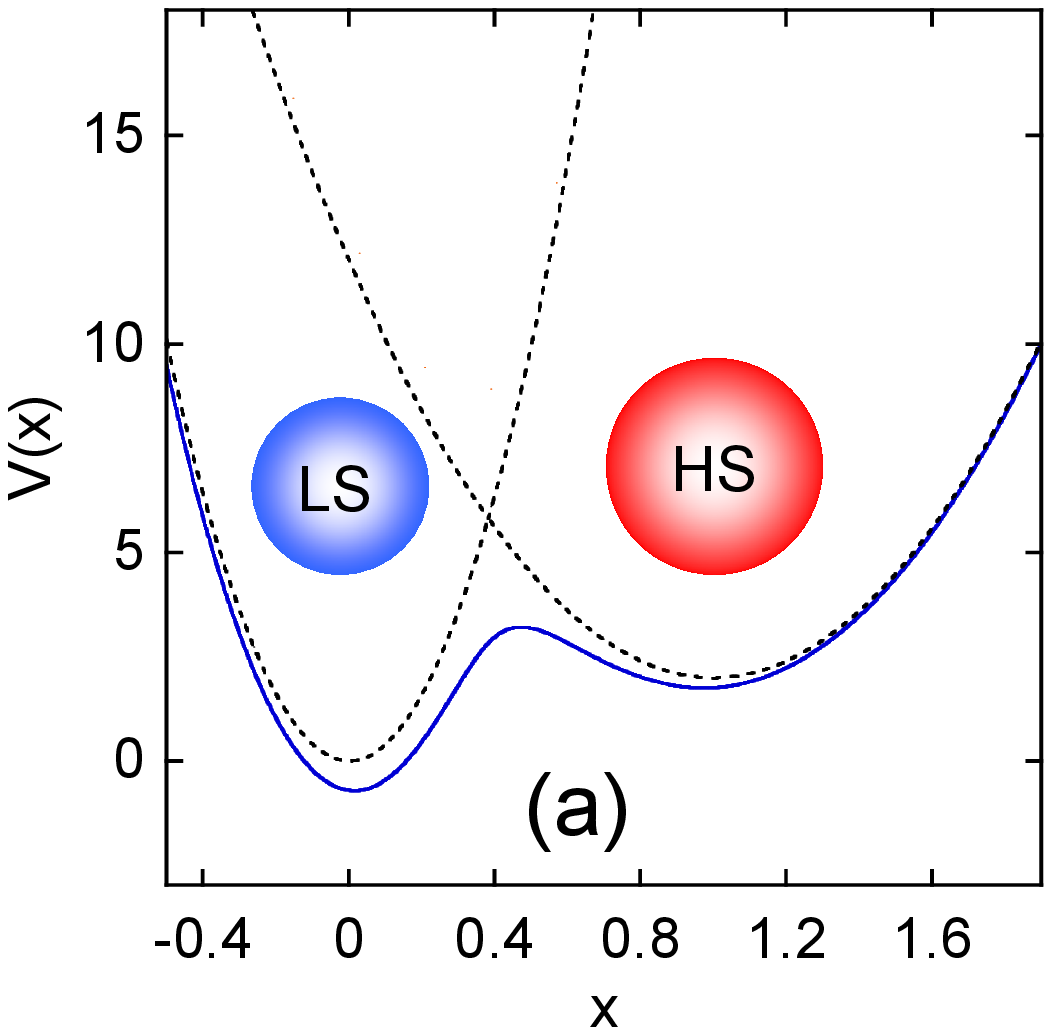}
  \end{center}
 \end{minipage}
  \begin{minipage}{0.49\hsize}
  \begin{center}
     \includegraphics[width=43mm]{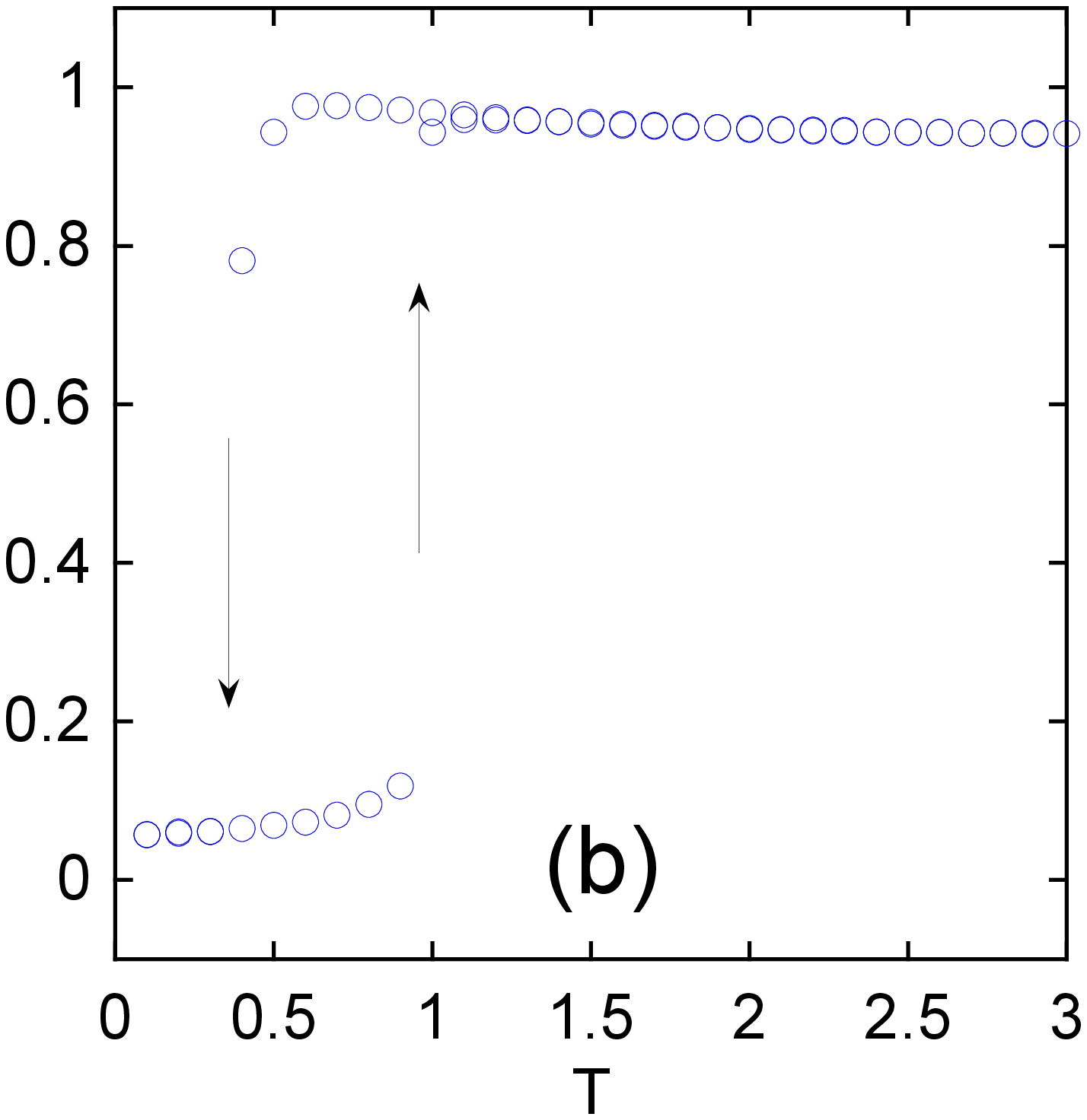}
  \end{center}
 \end{minipage}
\caption{(color online) (a) Intramolecular potential energy $V(x)$ shown by the
 solid (blue) curve. A realistic value $\frac{\omega_{\rm LS}}{\omega_{\rm HS}}=2$ is adopted \cite{Bousseksou3}. 
 The dotted curves are LS and HS potential energies
  without quantum mixing. A LS molecule (blue small circle) and HS molecule (red 
  large circle) are inset. 
 (b) Thermal hysteresis loop for $D=20$ and $L=100$ with the open boundary. 
 Open circles denote HS fraction.  
}
\label{Fig_model}
\end{figure}

We adopt a simple SC model for the square lattice, which represents general characteristics. 
In the model, both intramolecular and intermolecular interactions are taken into account \cite{Nishino_elastic2},
\begin{eqnarray}
{\cal H}_{0} = && \sum_{i=1}^{N}
\frac{ \mbox{ \boldmath $P$}_i^2}{2 M}  + \sum_{i=1}^{N} \frac{p_i^2}{2 m}
+ \sum_{i=1}^{N} V_i^{\rm intra}(r_i)    \\ 
&& +  \sum_{\langle i,j\rangle
} V_{ij}^{\rm inter}( {\mbox{ \boldmath $X$}_i},{\mbox{ \boldmath $X$}_
j},r_i,r_j).   \nonumber
\end{eqnarray}
Here, $\mbox{\boldmath $X$}_i$ and $\mbox{ \boldmath $P$}_i$ represent the coordinate and its conjugate momentum of the center of mass for the $i$th molecule. Conjugate variables $r_i$ and $p_i$ are defined for the totally symmetric mode 
for the $i$th molecule, which is the most important intramolecular motion 
\cite{Bousseksou3}. 
We also define the variable $x$ as $x=r-r_{\rm LS}$, where $r_{\rm LS}$(= 9) is the ideal radius of the LS molecule. That of the HS molecule is $r_{\rm HS}=r_{\rm LS}+1$. The intramolecular potential energy $V_i^{\rm intra}(x_i)$ is shown 
by the solid curve in Fig.~\ref{Fig_model} (a). 
We adopt the intermolecular potential
$V_{ij}^{\rm inter}( {\mbox{ \boldmath $X$}_i},{\mbox{ \boldmath $X$}_j},r_i,r_j)$ between the nearest neighbors and next-nearest neighbors,   
where $D$ is the strength of the intermolecular interaction \cite{Vinter}.

We study the present model by a molecular dynamics method,
in which we introduce a mechanism to control the large entropy difference between the HS and LS states \cite{Nishino_elastic2}. 
When the intermolecular interaction is stronger, the system exhibits a thermal hysteresis. Using PBC, uniform configurations are maintained during the transition between the HS and LS phases.  Using OBC, however, a macroscopic inhomogeneity is produced.

Here we focus on the dynamics of a relatively large hysteresis loop ($D=$20). 
In Fig.~\ref{Fig_model} (b) the temperature dependence of the 
HS fraction \cite{Nishino_elastic2} is 
given for a system of $N=L^2=100 \times 100$. 
The system was heated from $T=0.1$ to 3.0 in steps of 0.1, and then cooled 
to the initial temperature $T=0.1$. 
At each temperature, first 400,00 MD steps were discarded as transient time, 
and then 200,00 MD steps were used to measure physical quantities with the MD time step $\Delta t=0.01$.
The transition from the LS to HS state (LS $\rightarrow$ HS) occurs around $T=1.0$ in the heating process and from the HS to LS state (HS $\rightarrow$ LS) 
around $T=0.4$ in the cooling process.

\begin{figure}
\centerline{\includegraphics[clip,width=8.2cm]{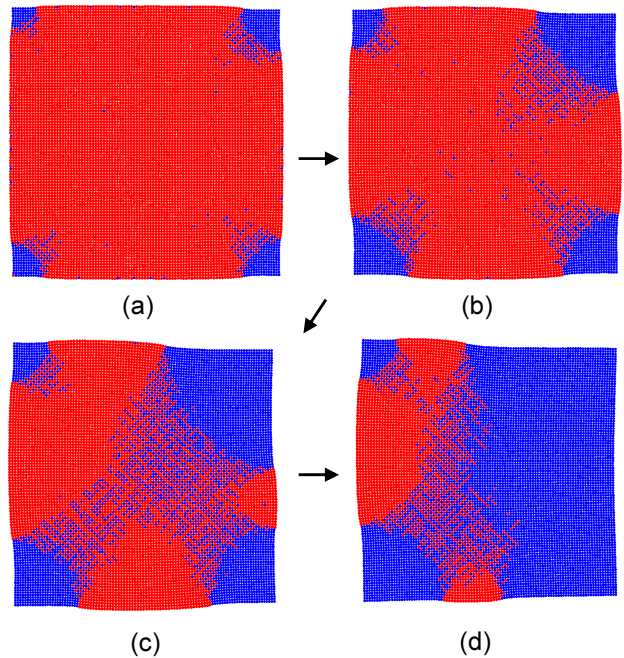} 
}
\caption{ (color online) Snapshots of configurations in the cooling process 
from the HS to LS phase. Red (blue) circles denote HS (LS) molecules. 
(a) $T=0.4$, $t=33108$, (b) $T=0.4$, $t=33506$, (c) $T=0.3$, $t=33692$, 
(d) $T=0.3$, $t=33804$. 
}
\label{Fig_HtoL}
\end{figure}

We clearly observed domain growth from corners 
during the transition from the HS to LS phase (Fig.~\ref{Fig_HtoL} (a)-(d)). 
LS domains grow and finally combine together to extend to the whole system. 
Domain growth occurs in the diagonal directions. 
We studied the system-size dependence of the process and found that clusters always grow from the corners, not from the sides (edges) 
or the inner part (bulk). 
The qualitatively same characteristics of clustering of LS domains were observed in relaxation processes from the metastable HS phase at a low temperature ($T
 < 0.4$), which is realized by LIESST in experiments.

The feature of clustering presents a distinct contrast to the cases of Ising-like models with short-range interactions, where the nucleation occurs from the inner part or the sides in large systems \cite{Rikvold}. 
In order to clarify this difference, we study energy dependence on the cluster pattern, and we find that the growth from sides is not acceptable in the elastic model. 

As the initial state we set a round LS domain (closed quadrant) 
at a corner in the complete HS phase. 
Then, we move all molecules slowly by reducing the total potential energy 
of the system. 
We define $N_{\rm LS}^{\rm I}$ as the number of LS molecules 
in the LS domain in the initial state, and 
$N_{\rm LS}^{\rm S}$ as that in the stationary state. 
$R_{\rm LS}^{\rm I}$ is the number of LS molecules 
in the horizontal ($X$) (or vertical ($Y$)) direction 
for the domain with $N_{\rm LS}^{\rm I}$.  

\begin{figure}[t]
\centerline{\includegraphics[clip,width=6cm]{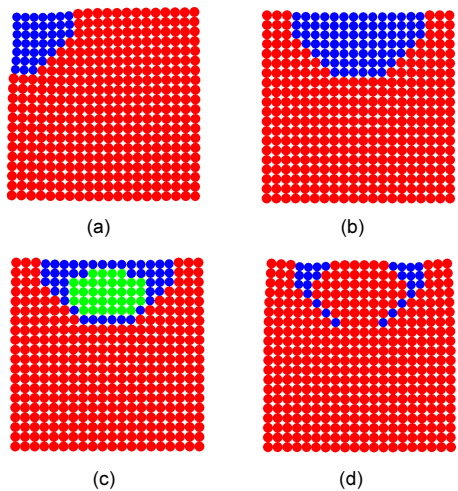}
}
\caption{(color online) (a) Stationary configuration of a round LS domain at 
a corner in HS phase. 
(b) Initial configuration of a round LS domain at a side 
(c) A configuration in the intermediate state. 
Green color denotes unstable state of molecules.  
(d) Configuration in the stationary state. 
Unstable LS molecules changed to HS molecules and 
the initially round LS domain does not exist any more. 
}
\label{equribrium_domains}
\end{figure}

Figure~\ref{equribrium_domains} (a) illustrates 
the configuration in the stationary state when we set  
$R_{\rm LS}^{\rm I}=7$ and $N_{\rm LS}^{\rm I}=39$. 
In this case, the quadrant shape was maintained 
and the number of LS molecules in the domain did not change 
during the simulation, i.e., $N_{\rm LS}^{\rm I}=N_{\rm LS}^{\rm S}$, 
although a small shift of the position and radius for each 
LS molecule was observed, which means that 
the LS domain in the stationary state is at least locally stable. 

\begin{figure}[t]
\centerline{
\includegraphics[clip,width=5.5cm]{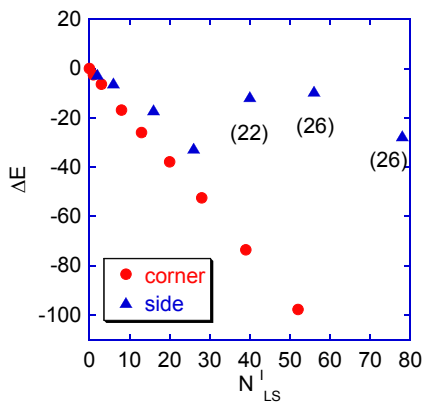} }
\caption{(color online) 
$\Delta E$ vs. $N_{\rm LS}^{\rm I}$ for LS domains at the corner and the side. 
For LS domains at the corner, $N_{\rm LS}^{\rm I}$ is the same as $N_{\rm LS}^{\rm S}$, unlike LS domains at the side. 
The number in a parenthesis is the value of $N_{\rm LS}^{\rm S}$ 
when the initial LS domain shape is not maintained and 
$N_{\rm LS}^{\rm S}$ varies from $N_{\rm LS}^{\rm I}$. }
\label{delatE_NLS}
\end{figure}

We investigate $\Delta E$, defined as the difference of energies 
between the stationary state and the complete HS phase. 
The dependence of $\Delta E$ on $N_{\rm LS}^{\rm I}$ is plotted by circles in Fig.~\ref{delatE_NLS}. 
We find that $\Delta E$ becomes lower as $N_{\rm LS}^{\rm I}$ increases, which indicates that the larger LS domain at the corner is energetically favorable. We checked that it holds true for larger system size ($L$).
Thus, the domain will grow if some small noise 
(thermal fluctuation) assists the system to relax. 

Next, we study the stability of LS domains (closed semicircle) at a side. 
Figure~\ref{equribrium_domains} (b) shows the initial state of 
the configuration of a LS domain ($R_{\rm LS}^{\rm I}=7$, 
$N_{\rm LS}^{\rm I}=78$) at a side. 
In this case, the initial configuration is found to be unstable due to high distortion. Several LS molecules change back to HS molecules and the number of LS molecules changes ($N_{\rm LS}^{\rm I} > N_{\rm LS}^{\rm S}$). 

Figure~\ref{equribrium_domains} (c) is a configuration in the 
intermediate state and the final stable configuration is shown in Fig.~\ref{equribrium_domains} (d). 
The molecules colored green (in gray in black-and-white print) 
in Fig.~\ref{equribrium_domains} 
(c) have intermediate radii (0.3 $< r <$ 0.7) between radii of LS and HS states. 
They feel high stress and the spin states of them are changing. 

In Fig.~\ref{delatE_NLS}, we plot as triangles 
$\Delta E$ for LS domains at the side as a function of $N_{\rm LS}^{\rm I}$. 
Unlike the case of the corner, $\Delta E$ is not a simple decreasing function of $N_{\rm LS}^{\rm I}$ for larger $N_{\rm LS}^{\rm I}$. 
It is worth noting that $N_{\rm LS}^{\rm S}$ is no more equal to 
$N_{\rm LS}^{\rm I}$ for $40 \le N_{\rm LS}^{\rm I}$, and takes 
a value for the configuration in the stationary state, which is given in a parenthesis in Fig.~\ref{delatE_NLS}.  LS domains which are bigger than a critical size are unstable at the side. 

If we set a round LS domain in the center of the HS phase, 
due to a huge distortion, the domain becomes very unstable 
and it collapses easily to reduce the number of LS molecules. 

We checked the qualitatively same tendency for the stability of LS domains 
when the system size ($L$) is larger. 
These observations lead to a major conclusion; LS domains cannot grow from the sides or the inner part of the system, which is different from the results of short-range interaction models. 
 
This conclusion suggests that different nucleation processes exist between short-range interaction models and the elastic model. 
We define $P_1$ (corner), $P_2$ (side), and $P_3$ (inner part) 
as the nucleation rate from a corner, a side, and the inner part, respectively. 
Generally, they are functions of the size and temperature, and 
$P_1 > P_2 > P_3$ because the surface energy increases in this order. 
However, if we take into account the possible location of nucleation, the probabilities to observe nucleation at a corner, a side, and the inner part of the system are $P_1$, $L \times P_2$, and $L^2 \times P_3$, respectively.  
Then, in short-range models, the relation $P_1 < L \times P_2 < L^2 \times P_3$ will hold for larger $L$ (linear dimension), and nucleation occurs in the inner part in large systems. 
Thus, so called multi-nucleation process takes place \cite{Rikvold}.

In the elastic model, however, $P_2$ and $P_3$ are essentially zero 
and $P_1$ (corner) is the only probability for nucleation.  
Therefore, even if the system size is large, nucleation (clustering) always starts from corners.

\begin{figure}
\centerline{\includegraphics[clip,width=8.2cm]{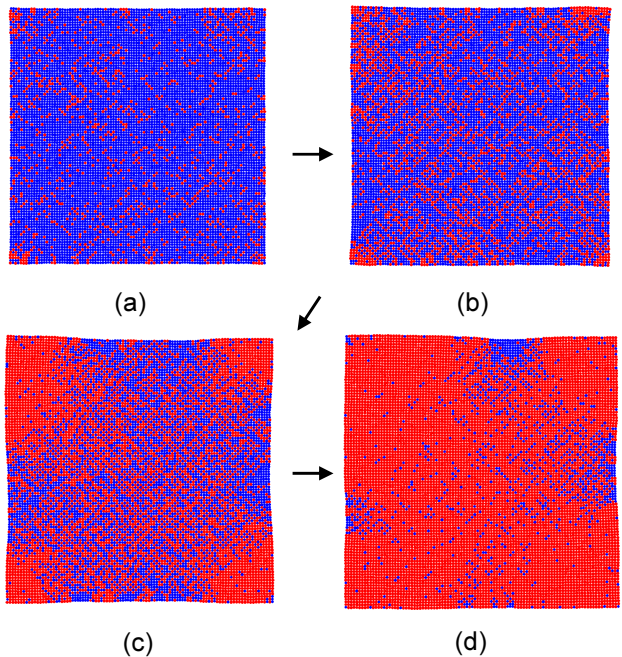} 
}
\caption{ (color online) Snapshots of configurations in the heating process from the LS to HS phase. Red (blue) circles denote HS (LS) molecules. 
(a) $T=1.0$, $t=5452$, (b) $T=1.0$, $t=5522$, (c) $T=1.0$, $t=5672$, (d) $T=1.0$, $t=5748$.  
}
\label{Fig_T_dep}
\end{figure}

We next investigate the process in heating. 
Snapshots of transient states from the LS to HS phase in the heating are given in Fig.~\ref{Fig_T_dep} (a)$-$(d). 
Here, we also find local clusters of HS molecules around the corners, but 
in contrast to the case of the process from the HS to LS phase (left branch of the hysteresis loop), a large homogeneous region appears as 
is observed in periodic boundary conditions. 

We consider the reason for the difference of the changing pattern 
between the heating (LS to HS) and cooling (HS to LS) processes. 
In the SC system, the HS and LS states are not equivalent and we may expect different types of relaxations for the cooling and heating processes. 
In the cooling process at low temperatures, the energy stability is more important than the entropy gain and the nucleation from a corner is the most favorable. In the heating process at high temperatures, however, 
the entropy gain becomes more important, and the configuration may change 
uniformly, which can be seen in the inner part of the system. 


In summary, we have studied effects of the boundary condition 
in a SC model with effective long-range interactions. 
We found that domains always grow from corners, which exhibits a striking contrast to the cases of short-range interaction models. 
In the heating process, an entropy-driven mechanism causes a smearing of 
clusters, and the configuration is close to that with the periodic boundary condition. 

The existence of macroscopic domains in SC compounds has been suggested in experimental studies of X-ray diffraction \cite{Huby,Pillet,Ichiyanagi}. 
The present study could give an insight into that suggestion. 
Dynamical properties of SC materials with OBC are important for studies of nano-scale systems, where the boundary plays a crucial role.   

The present work was supported by Grant-in-Aid for Scientific Research on Priority Areas (17071011) and for Scientific Research C (20550133), 
and by the Next Generation Super Computer Project, 
Nanoscience Program from MEXT of Japan. 
CE thanks to PNII 1994 Romanian CNCSIS Ideas Grant. 
The numerical calculations were supported by the supercomputer center of
ISSP of Tokyo University.


\begin{thebibliography}{99}
\bibitem{Vinter} 
The intermolecular potential \cite{Nishino_elastic} is defined 
as \\ 
$V_{ij}^{\rm inter}( {\mbox{ \boldmath $X$}_i},{\mbox{ \boldmath $X$}_j},
r_i,r_j)=f(d_{ij}-\Delta r)$, where $f(u) = D \left( e^{a' (u-u_0)} + e^{-b' (u-u_0)} \right)$. The variable $u_0$ is a constant such that $f(u)$ has the minimum at $u=0$ and $d_{ij}=| {\mbox{ \boldmath $X$}_i} - {\mbox{ \boldmath $X$}_j} |- (r_i+r_j)$. For the nearest neighbors, $\Delta r=0$, $a'=0.5$ and $b'=1.0$ are 
set, and for the next nearest neighbors, $\Delta r= 2(\sqrt{2}-1) \bar{r}$ with $\bar{r}=(r_{\rm LS} + r_{\rm HS}) /2$, $a'=0.1$ and $b'=0.2$.
\end{thebibliography}
\end{document}